%% file: TNOSP.tex
\documentclass[a4paper,fleqn]{cas-sc}

\usepackage{soul} 
\usepackage{color, xcolor} 

\usepackage[numbers]{natbib}
\usepackage[boxed,commentsnumbered,ruled,linesnumbered]{algorithm2e}

\newtheorem{definition}{Definition}  

\begin{document}
\shorttitle{Towards Top-$K$ Non-Overlapping Sequential Patterns} 
\shortauthors{Zefeng Chen et al.}


\title [mode = title]{Towards Top-\textit{K} Non-Overlapping Sequential Patterns}              



\author[1]{Zefeng Chen}
\ead{czf1027@gmail.com}
\address[1]{College of Cyber Security, Jinan University, Guangzhou 510632, China}

\author[1,2]{Wensheng Gan}
\ead{wsgan001@gmail.com}
\address[2]{School of Intelligent Systems Science and Engineering, Jinan University, Zhuhai 519070, China}

\author[1]{Gengsen Huang}
\ead{hgengsen@gmail.com}

\author[3]{Yan Li}
\ead{lywuc@163.com}
\address[3]{Hebei University of Technology, Tianjin 300131, China}

\author[4]{Zhenlian Qi}
\cormark[1]
\ead{qzlhit@foxmail.com}
\cortext[cor1]{Corresponding author}
\address[4]{Guangdong Eco-Engineering Polytechnic, Guangzhou 510520, China}


\begin{abstract}
    Sequential pattern mining (SPM) has excellent prospects and application spaces and has been widely used in different fields. The non-overlapping SPM, as one of the data mining techniques, has been used to discover patterns that have requirements for gap constraints in some specific mining tasks, such as bio-data mining. And for the non-overlapping sequential patterns with gap constraints, the Nettree structure has been proposed to efficiently compute the support of the patterns. For pattern mining, users usually need to consider the threshold of minimum support (\textit{minsup}). This is especially difficult in the case of large databases. Although some existing algorithms can mine the top-$k$ patterns, they are approximate algorithms with fixed lengths. In this paper, a precise algorithm for mining \underline{T}op-$k$ \underline{N}on-\underline{O}verlapping \underline{S}equential \underline{P}atterns (TNOSP) is proposed. The top-$k$ solution of SPM is an effective way to discover the most frequent non-overlapping sequential patterns without having to set the \textit{minsup}. As a novel pattern mining algorithm, TNOSP can precisely search the top-$k$ patterns of non-overlapping sequences with different gap constraints. We further propose a pruning strategy named \underline{Q}ueue \underline{M}eta \underline{S}et \underline{P}runing (QMSP) to improve TNOSP's performance. TNOSP can reduce redundancy in non-overlapping sequential mining and has better performance in mining precise non-overlapping sequential patterns. The experimental results and comparisons on several datasets have shown that TNOSP outperformed the existing algorithms in terms of precision, efficiency, and scalability.
\end{abstract}


\maketitle

\input{1_introduction.tex}
\input{2_relatedwork.tex}
\input{3_preliminaries.tex}
\input{4_technic.tex}
\input{5_experiment.tex}
\input{6_conclusion.tex}

\printcredits

\bibliographystyle{cas-model2-names}

\bibliography{TNOSP.bib}

\end{document}

%% file: 1_introduction.tex
\section{Introduction}

Rich data has become an indispensable part of everyone's lives in recent years. It has become an important topic to extract valuable information from various sequence data using sequential pattern mining (SPM) \cite{masseglia2005sequential,pinto2001multi}. Through SPM, users can discover and analyze statistically significant patterns from sequential items or event sequences \cite{kumar2019review}. In summary, SPM has a wide range of applications in various directions, such as web mining \cite{boggan2007gpus}, protein detection \cite{ferreira2005protein}, indoor trajectory \cite{chen2019indoor}, and so on. Especially when users need to discover some bio-data, such as genomes and protein sequences, the SPM with gap constraints is more meaningful than the general SPM \cite{li2012efficient}. Nevertheless, in these fields, gap constraints are required to be predefined. Hence, to face these mining scenarios and cope with various interesting data mining tasks, gap-constrained SPM algorithms have been proposed to minimize the number of unrelated and useless patterns for users \cite{orlando2004new,zhang2007mining}. For example, Antunes \textit{et al.} \cite{antunes2003generalization} put forward the problem of SPM with gap constraints, and Li \textit{et al.} \cite{li2008efficiently} proposed a method to efficiently mine closed subsequences with gap constraints. Pei \textit{et al.} \cite{pei2007constraint} studied the pattern-growth methods of constraint-based SPM to propose an extended framework. The most important advantage of these algorithms is that gap constraints allow users to meet their needs by setting gap flexibility, whereas it is difficult to solve the problem of sequential pattern mining with gap constraints. For an Apriori-like mining algorithm, all super-patterns of an infrequent pattern are infrequent, but all sub-patterns of a frequent pattern are frequent. It provides a powerful mining pruning strategy, i.e., in an SPM task, if the support of a mined pattern is less than the minimum support threshold, all super-patterns of this pattern can be removed from the candidate set. In such a case, the non-overlapping condition needs to be applied. It is more decided than the no-condition for pattern mining and less restrictive than the one-off condition. Consequently, Wu \textit{et al.} \cite{wu2017nosep} proposed a method of non-overlapping SPM with gap constraints named NOSEP, where non-overlapping sequential patterns can be discovered by the Apriori property. Note that an Apriori-like SPM algorithm can calculate the support of the patterns that have high importance in a level-wise way. Thus, a large number of the patterns will be filtered from consideration if their sub-patterns are not frequent patterns.

Finding all frequent sequences in large databases is a difficult task \cite{zaki2001spade}. In addition, although the algorithms mentioned above can be used to search patterns, it is still a tough problem to set the minimum support threshold (\textit{minsup}) without prior knowledge \cite{chai2018top,zhang2021tkus}. The greater the \textit{minsup}, the fewer frequent patterns can be discovered. If users set a high \textit{minsup}, some potential desired patterns may be missed. As \textit{minsup} decreases, more patterns will be mined as frequent patterns. It may require a longer running time and lead to a certain waste of resources. Therefore, finding frequent patterns effectively and discovering valuable patterns has become a vital topic \cite{chai2018top}. To address this challenge, top-$k$-based algorithms provide a workaround to change the mining task to discover top-$k$ frequent patterns. The notation $k$ refers to the number of patterns in processed datasets \cite{tzvetkov2005tsp,wang2005tfp,wu2012mining}. Top-$k$, literally, is the $k$ most frequent patterns. Generally, top-$k$ pattern mining can enormously reduce the number of excessive patterns and effectively reserve patterns with high frequency when using the frequent pattern mining algorithms \cite{chai2018top}. As mentioned in NOSEP \cite{wu2017nosep}, SPM with the non-overlapping condition can compensate for the disadvantages of the one-off condition and the no-condition. It has outstanding foundations and prospects in SPM. NOSEP mines all the sequential patterns whose support is larger than the setting \textit{minsup}. 

Utilizing the top-$k$-based algorithm, users can directly set the number of patterns $k$ instead of considering the minimum support threshold \textit{minsup}. For non-overlapping sequential pattern mining tasks, using $k$ to discover patterns is more efficient than using \textit{minsup} in some application scenarios. For example, the distribution of CpG islands\footnote{https://en.wikipedia.org/wiki/CpG\_island\_hypermethylation} (dinucleotides located in human gene promoters) is very heterogeneous. In certain segments of the genome, CpG islands remain at or above their normal probability \cite{wang2014efficient}. The gap constraints and length constraints (their definitions will be presented later in Definitions \ref{definition 3} and \ref{definition 4} in Section \ref{sec:preliminaries}) need to be set. When mining frequent sequential patterns, the gap constraint means that there are several other items between the two items in the patterns that can be mined. Some significant gaps in sequential patterns whose items are not continuous can thus be avoided. The related research was reported to discover the CpG island dinucleotides in the human genome and compare them in different segments of the genome. These may help to better understand the changes of CpG island methylation in human development, aging, and disease \cite{bock2006cpg}. In general, we want to conveniently mine the top several patterns of a certain length. Obviously, it is more intuitive to find valuable rules for CpG islands with the number of $k$ patterns in DNA sequences. It makes the sequence detection of CpG islands more convenient. As a result, not only can the SPM with gap constraints be effectively used, but also the patterns corresponding to the top-$k$ of the required support can be obtained directly. Similarly, this mining algorithm can also be used in other fields. Chai \textit{et al.} \cite{chai2018top} proposed NOSTOPK to mine the top-$k$ sequential patterns with gap constraint under non-overlapping conditions. However, it should be noted that the mining task is to discover the top-$k$ patterns under each length in NOSTOPK. It is different from that of traditional top-$k$ sequence pattern discovery.

In this study, a top-$k$ SPM mining algorithm based on Nettree is designed, named \underline{T}op-$k$ \underline{N}on-\underline{O}verlapping \underline{S}equential \underline{P}attern mining (TNOSP). After unifying the indicators, the TNOSP algorithm is more precise and has better properties than the heuristic-based NOSTOPK algorithm. There is only one algorithm proposed to discover top-$k$ patterns in non-overlapping SPM, and its results are approximate. However, it only works for fixed-length patterns and cannot discover the most useful patterns in some cases. The major contributions of this paper are as follows:

\begin{itemize}
    \item We define the related concepts of top-$k$ non-overlapping sequential pattern and formalize the problem of top-$k$ non-overlapping SPM. Using the top-$k$ strategy is more practical than the traditional setting of the minimum support threshold in data mining.

    \item The paper proposes TNOSP, an effective and efficient algorithm for mining useful patterns without setting a minimum support threshold and reducing the number of excessive patterns. By using a \underline{Q}ueue \underline{M}eta \underline{S}et \underline{P}runing strategy named QMSP, the TNOSP algorithm can significantly reduce running time and computation costs.

    \item A series of experiments have been conducted on the selected real DNA datasets to verify the high efficiency and precision of the TNOSP algorithm compared with the algorithm that is designed to mine the top-$k$ non-overlapping sequential patterns.
\end{itemize}

The other chapters of this paper cover the following. In Section \ref{sec:relatedwork}, previous studies on SPM, non-overlapping SPM, and top-$k$ pattern mining are briefly reviewed. Section \ref{sec:preliminaries} introduces the definition of top-$k$ non-overlapping SPM. The efficient algorithm namely TNOSP is proposed in detail in Section \ref{sec:algorithm}. The experimental evaluations are presented in Section \ref{sec:experiments}. In Section \ref{sec:conclusion}, we draw the conclusions of this paper and discuss future work.

%% file: 2_relatedwork.tex
\section{Related Work} \label{sec:relatedwork}

In this section, we discuss sequential pattern mining (SPM), non-overlapping sequential pattern mining, and top-$k$ pattern mining, respectively.

\subsection{Sequential Pattern Mining} 

One of the most classic tasks of data mining is discovering useful patterns in sequences \cite{fournier2017survey}. Through the algorithm of SPM, the user inputs the sequence database, and all sequences with their supports that are not less than the minimum support threshold can be output. The reason why SPM is so popular is that we can discover patterns in large databases and explain these data through mining tasks. These patterns mined by SPM are helpful for understanding the data and making decisions. Therefore, SPM has many real-life applications in many fields. It is a very active research topic where new algorithms and applications have been presented in a large number of papers recently, including many extensions of SPM for different needs in various fields. Agrawal and Srikant \cite{agrawal1995mining} came up with the SPM in 1995, mining sequential patterns. Since then, the topic of SPM has gained more attention. The first simple algorithm for SPM is AprioriAll, which is based on the Apriori property \cite{agrawal1994fast}. It is covered in the previous section. Afterward, many algorithms of SPM have been proposed \cite{fournier2017survey,gan2019survey,mabroukeh2010taxonomy}. For example, Zaki \textit{et al.} \cite{zaki2001spade} proposed SPADE to reduce the time of scanning a database by the method of vertical database representation. Ayres \textit{et al.} \cite{ayres2002sequential} proposed SPAM to reduce the costs of inputs and outputs with a depth-first search strategy, and Han \textit{et al.} \cite{han2001prefixspan}  proposed PrefixSpan to make pattern grow quickly by using two projection strategies. However, algorithms will generate many candidates while handling dense datasets resulting from the combinatorial explosion. Accordingly, Yang \textit{et al.} \cite{yang2007lapin} developed LAPIN to solve this problem straightforwardly. In other words, the last occurrence of an item determines whether to continue concatenating candidates for that item to generate super-candidates for that item. In addition, with the increasing application of SPM, more interesting and useful discoveries of SPM have been researched for improving mining efficiency, such as parallel SPM \cite{gan2019survey}, high-utility sequential pattern mining \cite{gan2019proum,gan2021explainable}, significance-based discriminative SPM \cite{he2019significance}, constraint-based SPM \cite{van2018mining,huynh2022efficient}, and so on. Some applications require algorithms that allow for a certain gap between two itemsets. However, there is a common drawback in these algorithms: the gap constraint is not considered, which cannot meet the needs of mining tasks under some specific mining tasks.

\subsection{Non-overlapping Sequential Pattern Mining}

The drawbacks of the above algorithms have been discussed before. Antunes \textit{et al.} \cite{antunes2003generalization} first introduced the concept of gap-constraint SPM. Ji \textit{et al.} \cite{ji2007mining} proposed minimal sub-sequence mining with gap constraints. Non-overlapping SPM is a branch of gap-constraint SPM, which means that no same sequence letter can appear twice in the same position \cite{wu2020netncsp}. In this section, non-overlapping sequential pattern mining is discussed. A letter may match one or more pattern letters, and under the matching of multiple letters, different matches may lead to different frequency count results. To meet these challenges, the existing algorithms generally have three conditions to calculate the frequency of patterns, which are the no-condition \cite{zhu2007mining}, the one-off condition \cite{lam2014mining}, and the non-overlapping condition \cite{ding2009efficient}, respectively. Mining tasks with no-condition don't satisfy the Apriori property, such as the supports of $A[0, 3]G$ and $A[0, 3]G[0, 3]C$ in \textit{AAGGGCC} are respectively 6 and 12. However, mining tasks with the condition of being one-off and non-overlapping meet the Apriori property. In addition, the non-overlapping condition is more detailed than the no-condition but less limited than the one-off condition. Therefore, SPM with the non-overlapping condition has a higher probability of finding more meaningful patterns from the sequence or dataset in SPM \cite{wu2017nosep}. In recent years, numerous studies on the topic of non-overlapping SPM have been conducted. For example, NTP-Miner \cite{wu2021ntp} is studied to adopt strategies of depth-first and backtracking to improve time efficiency as well as space efficiency. NetNPG \cite{shi2020netnpg} addresses a non-overlapping SPM problem with general gaps, indicating that the gap can be not only natural numbers but negative values. Meanwhile, HANP-Miner \cite{wu2021hanp} used a simplified Nettree, which significantly reduced the algorithm's space and time complexity. The goal of this study is to design an algorithm that has a higher efficiency for mining sequential patterns with non-overlapping. NOSEP \cite{wu2017nosep} discovers sequential patterns by employing the Nettree, which is more effective than other algorithms \cite{wu2014mining}. Hence, the optimization and improvement of NOSEP will be described in this paper.

\subsection{Top-\textit{k} Pattern Mining} 

Though NOSEP is able to efficiently search patterns and calculate the support values, setting the minimum support threshold is subtle. Because a small threshold may result in a large number of patterns, which is computationally expensive, while a high threshold may result in some patterns that do not meet users' requirements. To address this challenge, top-$k$-based algorithms \cite{han2002mining,tzvetkov2005tsp} are used to provide users with an adequate opportunity to directly decide the required number of patterns rather than setting the minimum support threshold \cite{chai2018top}. Top-$k$ pattern mining, a classic strategy to enormously reduce the number of excessive patterns, can effectively reserve patterns with high-frequency \cite{chai2018top}. Han \textit{et al.} \cite{han2002mining} introduced mining frequent patterns using the $k$ frequent patterns, where $k$ is the number of top frequent patterns. Since then, this strategy has become popular in pattern mining. Tzvetkov \textit{et al.} \cite{tzvetkov2005tsp} proposed TSP to make use of the length constraint and the properties of top-$k$ closed sequential patterns to perform dynamic support raising and projected database pruning. Wang \textit{et al.} \cite{wang2005tfp} developed an efficient algorithm based on the FP-tree that can be pruned dynamically both during and after the construction of the FP-tree. Consequently, the TNOSP algorithm for top-$k$ SPM with the non-overlapping condition is proposed in this paper. Users can discover the non-overlapping SPM with gap constraints by employing the Nettree which has a significantly higher efficiency than other structures in SPM algorithms. TNOSP can also reduce the number of excessive patterns by utilizing top-$k$ pattern mining.

%% file: 3_preliminaries.tex
\section{Preliminary and Problem Statement} \label{sec:preliminaries}

In this section, basic concepts and notations related to non-overlapping sequential pattern mining are introduced. To more clearly explain the addressed problem, some definitions are adopted from previous studies that have been explained. What's more, the problem definition of TNOSP is formalized. Definitions related to sequences are given below.

\begin{table}[h]
    \centering
    \caption{Sequence database}
    \label{table1}
    \begin{tabular}{|c|c|}  
	\hline 
        \textbf{SID} & \textbf{Sequence} \\
	\hline  
	$S_{1}$ & $<$\textit{A}\textit{G}\textit{G}\textit{A}\textit{T}$>$ \\ 
	\hline
	$S_{2}$ & $<$\textit{A}\textit{T}\textit{G}\textit{G}$>$ \\ 
	\hline
	$S_{3}$ & $<$\textit{C}\textit{C}\textit{T}\textit{A}\textit{T}\textit{A}$>$ \\  
	\hline  
	$S_{4}$ & $<$\textit{G}\textit{C}\textit{G}\textit{C}\textit{G}\textit{T}$>$ \\  
	\hline  
    \end{tabular}
\end{table}

\begin{definition}[Sequence database]	
    \label{definition 1}
    \rm In a database, let a set of items $I$ = \{$i_{1}$, $i_{2}$, $i_{3}$, ..., $i_{N}$\}. As an itemset, $X$ has one or more items of $I$ as $X \subseteq I$, which is not empty. The notation $|X|$ is called the set cardinality and represents the number of items in the itemset $X$. When the itemset $X$ includes $k$ items, represented by $|X|$ = $k$ mathematically, it is stated to be an $k$-itemset with a length of $k$. However, in the task of non-overlapping sequential pattern mining, there is usually only one item in an itemset. In other words, $k$ = 1, and all itemsets are 1-itemsets. In a sequence like $S$ = $<$$X_{1}$ $X_{2}$...$X_{n}$$>$, $X_{i}$ are in front of $X_{i+1}$ where $i < n$, and their orders can't be changed. In a nutshell, all sets of items in $S$ are ordered, and $n$ represents the size of $S$. Besides, $L$-sequence is a sequence $S$ with the length of $L$, calculated by $L$ = $\sum_{k = 1}^{n}|X_{k}|$. In our task, $L$ = $n$. In addition, $\mathcal{D}$ represents a sequence database, which is denoted as  $\mathcal{D}$ = \{$S_{1}$, $S_{2}$, $S_{3}$, ..., $S_{n}$\}. There are a set of tuples $\left<\textit{SID}, \textit{DS}\right>$ in each $\mathcal{D}$, where \textit{DS} is the data sequence and \textit{SID} is the unique identifier for \textit{DS}.	
\end{definition}

\begin{definition}[Support of patterns]
    \label{definition 2}
    \rm The support of itemset $X$ indicates the frequency of the itemset $X$ in a database, with a symbol of \textit{sup}$\left(X\right)$. For example, given a set of transactions $\mathcal{D}$, the goal of mining sequential patterns is to find all sequential patterns whose support is not less than the minimum support threshold, denoted as a symbol of \textit{minsup}. In traditional SPM algorithms, the \textit{minsup} is set by users. Patterns with a support value less than \textit{minsup} will not appear frequently in the result. In the sequence database $\mathcal{D}$, the pattern $P$'s support is the sum of the supports of $P$ in each sequence $S$, calculated as \textit{sup}($P$, $\mathcal{D}$) = $\sum_{i = 1}^{n}|\textit{sup}(P,S_{i})|$.
\end{definition}

For instance, in Table \ref{table1}, \textit{sup(A, $S_{1}$)} = 2, \textit{sup(A, $S_{2}$)} = 1, \textit{sup(A, $S_{3}$)} = 2, \textit{sup(A, $S_{4}$)} = 0. Thus, in $\mathcal{D}$, \textit{sup(A)} = 2 + 1 + 2 + 0 = 5, \textit{sup(A)} = 5, \textit{sup(C)} = 4, \textit{sup(G)} = 7, \textit{sup(T)} = 5. If \textit{minsup} = 5, then \textit{C} and its super-patterns will be pruned according to the Apriori property \cite{agrawal1994fast}.

\begin{definition}[Gap constraint]
    \label{definition 3}
    \rm The mined pattern $P$ with gap constraints is represented as $p_{1}$ $\left[min_{1}, max_{1}\right]$ $p_{2}$ ... $\left[min_{m-1}, max_{m-1}\right]$ $p_{m}$, where $p_{j}$ $\subseteq$ $S$, $j$ = 1, 2, 3, ..., $min_{j}$, and $max_{j}$ are two non-negative integers that represent the minimum and maximum gap constraint, respectively, denoted as \textit{gap} = $\left[\textit{mingap}, \textit{maxgap}\right]$. A gap constraint \textit{gap} = $\left[\textit{mingap}, \textit{maxgap}\right]$ means that $min_{1}$ = $min_{2}$ = $min_{3}$ = ... = $min_{n-1}$ = \textit{mingap} and $max_{1}$ = $max_{2}$ = $max_{3}$ = ... = $max_{n-1}$ = \textit{maxgap} in $P$. In this study, the gap constraints can be set by users according to their demands.
\end{definition}

For example, as shown in Table \ref{table1}, if \textit{gap} = [0,2], \textit{sup(AT, $S_{1}$)} = 1, \textit{sup(AT, $S_{2}$)} = 1, \textit{sup(AT, $S_{3}$)} = 1, \textit{sup(AT, $S_{4}$)} = 0, so in $\mathcal{D}$, \textit{sup(AT)} = 1 + 1 + 1 + 0 = 3. Considering $S_{4}$ = $s_{1}$$s_{2}$$s_{3}$$s_{4}$$s_{5}$$s_{6}$ = \textit{GCGCGT} and the pattern $P$ = \textit{G}[0,2]\textit{C}[0,2]\textit{G}, the occurrences of $P$ in $S_{4}$ $<$1, 2, 3$>$, $<$1, 2, 5$>$, $<$1, 4, 5$>$, and $<$3, 4, 5$>$ are all satisfy the conditions. Conversely, if \textit{gap} = [0,1], only $<$1, 2, 3$>$ and $<$3, 4, 5$>$ satisfy the conditions.

\begin{definition}[Length constraint]
    \label{definition 4}
    \rm The length constraint can be noted as \textit{len} = [\textit{minlen}, \textit{maxlen}], which means that the length constraint is between the minimum length constraint and the maximum length constraint. If the length $L$ of a sequence in a pattern satisfies \textit{minlen} $\le$ \textit{L} $\le$ \textit{maxlen}, it is considered to meet the length constraint.
\end{definition}

In Table \ref{table1}, the sequence $S_{4}$ = $s_{1}$$s_{2}$$s_{3}$$s_{4}$$s_{5}$$s_{6}$ = \textit{GCGCGT} and \textit{gap} = [0,2]. If \textit{len} = [0,5], the occurrences of $P$ in $S_{4}$ $<$1, 2, 3$>$, $<$1, 2, 5$>$, $<$1, 4, 5$>$, and $<$3, 4, 5$>$ are all satisfy the conditions. The length of the occurrences $<$1, 2, 5$>$ is then calculated as \textit{len} = 5 - 1 + 1 = 5. If \textit{len} = [1,4], only $<$1, 2, 3$>$ and $<$3, 4, 5$>$ meet the conditions.

\begin{definition}[Non-overlapping condition]
    \label{definition 5} 
    \rm If there are two occurrences $L_{1}$ = $<$$l_{1}$, $l_{2}$, $l_{3}$, ..., $l_{\mu}$$>$ and $L_{2}$ = $<$$l_{1}^{'}$, $l_{2}^{'}$, $l_{3}^{'}$, ..., $l_{\mu}^{'}$$>$, and in the case of ${\forall}$ $1 \le i$ $\le \mu$, it meets the condition of $l_{i}$ $\neq$ $l_{i}^{'}$, then $L_{1}$ and $L_{2}$ are called two non-overlapping occurrences. When all occurrences in an occurrence set are non-overlapping, the occurrence sets are defined as non-overlapping sets of occurrences. Accordingly, for $P$ in $S$, the support is the size of the maximum non-overlapping sets of occurrences.
\end{definition}

For example, in Table \ref{table1}, when \textit{gap} = [0, 2] and \textit{len} = [0, 5], the occurrences of $P$ in $S_{4}$ $<$1, 2, 3$>$, $<$1, 2, 5$>$, $<$1, 4, 5$>$, and $<$3, 4, 5$>$ all satisfy the conditions. However, the two sets of occurrences $<$1, 2, 3$>$ and $<$1, 2, 5$>$, $<$1, 4, 5$>$ and $<$3, 4, 5$>$ are overlapping. Accordingly, the support of $P$ in $S_{4}$ is 2 rather than 4, which is the size of the maximum non-overlapping set of occurrences.

\begin{definition}[Top-$k$ non-overlapping sequential pattern mining]
    \label{definition 6} 
    \rm In top-$k$ non-overlapping SPM, the frequent patterns are those that correspond to the top-$k$ supports in the mined sequence $S$ or mined sequence database $\mathcal{D}$ under the condition of gap and length constraints. The mining task in this paper is to search for all these frequent patterns under non-overlapping conditions. 
\end{definition}

%% file: 4_technic.tex
\section{The TNOSP Algorithm}  \label{sec:algorithm}

Based on the discussion mentioned above, this paper shows a novel algorithm named TNOSP for mining the top-$k$ patterns under the non-overlapping condition precisely and efficiently. This algorithm can be subdivided into three subsections for discussion. In Section \ref{NETGAP algorithm}, the algorithm based on Nettree for mining non-overlapping sequential patterns will be introduced. In Section \ref{top-k algorithm}, we develop the top-$k$ algorithm named TNOSP. In Section \ref{QMSP}, the queue meta set pruning (QMSP) strategy is proposed.

\subsection{Nettree-based Algorithm}
\label{NETGAP algorithm}

When calculating the support of patterns, a major problem that needs to be solved is that the occurrence is difficult to find without using a backtracking strategy. To solve the problem, the NETGAP algorithm was proposed by Wu \textit{et al.} \cite{wu2017nosep}. In Table \ref{table1}, the sequence $S_{1}$ = \{$s_{1}$$s_{2}$$s_{3}$$s_{4}$$s_{5}$\} = \textit{AGGAT}, and if users need to check the pattern $P_{1}$ = \textit{AG} with \textit{gap} = [0, 1], the first non-overlapping occurrence is <1, 2> and <1, 3> won't be found due to the non-overlapping condition. Users who search the pattern $P_{2}$ = \textit{AGT} will not find the occurrence <1, 2, 5> because 5 - 2 - 1 = 2 does not satisfy the condition: \textit{gap} = [0, 1]. The occurrence <1, 3, 5> is treated as non-overlapping occurrences of the pattern $P_{2}$ = $A$[0, 1]$G$[0, 1]$T$, which is one of the super-patterns of $P_{1}$. In some algorithms like INSgrow \cite{pan2009efficient}, if the occurrence of the pattern is calculated according to the sub-occurrences of its sub-patterns, some feasible occurrences may be lost without using the backtracking strategy, such as the <1, 3, 5> mentioned above \cite{ding2009efficient}. To solve this problem efficiently, the NETGAP algorithm without using a backtracking strategy was proposed.

\begin{definition}[Nettree structure]
   \label{definition 7}
    \rm Nettree \cite{wu2010nettree} is a tree-based structure, which includes the elements of \textit{parent}, \textit{child}, \textit{level}, \textit{root}, \textit{leaf}, and so on. Nettree may have \textit{roots} of more than one, which is different from the tree structure, which can only have one \textit{roots}. Accordingly, for easily describing a node, we define that the node $i$ in the $j^{th}$ \textit{level} is noted by $n^{i}_{j}$. Another feature that distinguishes it from tree structure is that every non-\textit{root} root node may have more than one \textit{parent}, but all of its \textit{parents} must be at the same \textit{level}. From a node to the \textit{root} node, there may be various paths. Responsively, a \textit{leaf} in the last \textit{level} of Nettree is defined as an absolute \textit{leaf}. Besides, in a Nettree, a full path means a path from a \textit{root} to an absolute \textit{leaf}.
\end{definition}

\begin{figure}[ht]
	\centering
	\includegraphics[clip,scale=0.32]{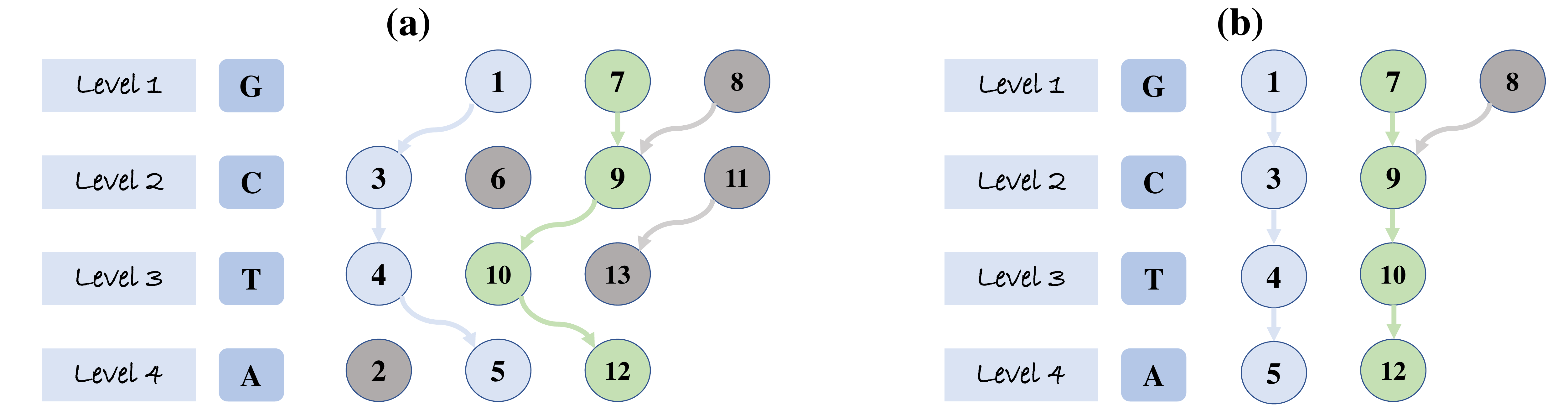}
	\caption{The Nettree for a pattern in a sequence.}
	\label{figure 1}
\end{figure}

Given a sequence $S$ = \{$s_{1}s_{2}s_{3}s_{4}s_{5}s_{6}s_{7}s_{8}s_{9}s_{10}s_{11}s_{12}s_{13}$\} = \textit{GACTACGGCTCAT}, a pattern $P$ = \{$p_{1}$$p_{2}$$p_{3}$$p_{4}$\} = \textit{GCTA}, the gap constraints \textit{gap} = [0, 2], and the length constraints \textit{len} = [1, 6]. According to $P$ and $S$, the algorithm can create its Nettree, as shown in Figure \ref{figure 1}(a).

In this paper, the task is to discover the patterns in the sequence, which can be divided into three parts. The pseudocode of the algorithm we propose is shown in Algorithm \ref{alg:NETGAP}.

\begin{itemize}	
    \item Firstly, the leaves of the occurrences of two adjacent \textit{levels} need to be connected if the occurrence of high-\textit{level} is higher than the occurrence of low-\textit{level} under the conditions of gap constraints and length constraints. For example, in Figure \ref{figure 1} (a), $n_{1}^{1}$ should be connected with $n_{2}^{3}$, $n_{2}^{3}$ should be connected with $n_{3}^{4}$, and $n_{3}^{4}$ should be connected with $n_{4}^{5}$ as well.
	
    \item Secondly, the leaves that have no path to reach the leaf of the previous \textit{level} (unless it is the first node) or the next \textit{level} (unless it is the last node) should be pruned, which is defined as a lonely node. For example, the node $n_{4}^{2}$ is pruned because it has no path to the previous \textit{level} and is not the first node. The $n_1^8$, $n_2^6$, $n_2^11$, and $n_313$ are also pruned in the same situation.
	
    \item Third, it is clear that after pruning, the first occurrence $<$1, 3, 4, 5$>$ is obtained in the updated Nettree, which is highlighted in light blue in Figure \ref{figure 1}. The acquired occurrences are pruned after the traversal is complete. Repeating this process, the occurrence of $<$7, 9, 10, 12$>$ is found. After finding the pattern, the corresponding leaf nodes are also pruned until there is no more pattern to discover, so the occurrences $<$1, 3, 4, 5$>$ and $<$7, 9, 10, 12$>$ are pruned in order. At this point, the excavation process is finished, and there are two non-overlapping occurrences, which are $<$1, 3, 4, 5$>$ and $<$7, 9, 10, 12$>$, for $P$ in $S$. \textit{sup(P, S)} = 2 is thus obtained.
\end{itemize}

\begin{algorithm}[h]
	\caption{NETGAP algorithm}
	\label{alg:NETGAP}
	\LinesNumbered
	\KwIn{sequence $S$; the mined pattern $P$; the miminum support threshold \textit{minsup}; gap constraint \textit{gap} = [\textit{mingap, maxgap}] and length constraint \textit{len} = [\textit{minlen, maxlen}].} 
	\KwOut{\textit{sup}($P$, $S$).}	
	Create a Nettree of $P$ to $S$\;
	Connect the leaves of occurrences of two adjacent \textit{levels} under the eligible condition\;
	Prune patterns at lonely node\;
	
	\For{$n_{1}^{j} \in $ \textit{Nettree} \do}{
		\textit{node}[1] $\leftarrow$ $n_{1}^{j}$;\qquad $//$ Occurrences are stored in nodes\\
		\For{$j=1$ to \textit{Nettree.level} -1 \do}{
			\textit{node}[$j$+1] = the leftmost \textit{child} of \textit{node}[$j$] satisfied the length constraints\;
		}	
		\textit{sup}($P$, $S$) $\leftarrow$ \textit{sup}($P$, $S$) + 1\;	
		prune the \textit{node}[$j$]\;
	}
	\textbf{return} \textit{sup}($P, S$)
	
\end{algorithm}

\subsection{The TNOSP Algorithm} \label{top-k algorithm}

Before introducing the top-$k$ algorithm, the \textit{generateCandidate} procedure is first introduced. This procedure \cite{wu2017nosep} generates the candidates. For example, the mining aims to discover all frequent patterns in the sequence $S$ = $s_1s_2s_3s_4s_5s_6s_7s_8s_9$ = \textit{AGTTAGCAC} with the conditions: \textit{minsup} = 2, gap constraints \textit{gap} = [0, 3], and length constraints \textit{len} = [1, 8]. After analysis, four types of frequent patterns with a length of 1 are \textit{A}, \textit{C}, \textit{G}, \textit{T}, and five types of frequent patterns with a length of 2 are {\textit{AA}, \textit{AC}, \textit{AG}, \textit{GA}, \textit{TA}}. Therefore, the number of candidate patterns with a length of 3 is 5 $\times$ 4 = 20, because four types of candidate patterns can be generated by each kind of frequent pattern with a length of 2. However, after analysis, the candidate set has such characteristics that a super-pattern with an infrequent pattern suffix will also not be a frequent pattern. Though \textit{AC} is a frequent pattern, the pattern \textit{ACG} is not frequent, because the pattern \textit{CG} is not frequent. Using the pattern growth approach, there are thirteen different types of candidate patterns with a length of three. They are respectively \textit{AAA}, \textit{AAC}, \textit{AAG}, \textit{ACA}, \textit{ACG}, \textit{AGA}, \textit{ATA}, \textit{GAA}, \textit{GAC}, \textit{GAG}, \textit{TAA}, \textit{TAC}, and \textit{TAG}, which will be put into the candidate set with length $L$ + 1. The procedure of \textit{generateCandidate} is shown in Algorithm \ref{alg:top-k}. On the assumption that a candidate pattern of length $L$ exists, it is used to generate candidate patterns with length $L$ + 1. This method of pattern growth is more efficient and significantly less computationally intensive; hence, this method continues to be adopted in this paper.
	
With the Nettree-based method and the \textit{generateCandidate} algorithm, algorithms for extracting top-$k$ sequential patterns with the non-overlapping condition have been developed \cite{chai2018top} named NOSTOPK. However, this algorithm is a heuristic algorithm; in other words, it is inaccurate and may miss some critical patterns. This algorithm is also used to find the top-$k$ patterns for any given length $L$. As a result, if users need to find top-$k$ patterns in all patterns and the length of the pattern is limited to less than $L$, NOSTOPK must specify the length $L$ of mining. The process won't stop unless it reaches the specified $L$. It is difficult for users to sum the patterns of all lengths and compute $k$. It is also difficult for users to set $L$ without specifying \textit{minsup}. As a result of these two challenges, TNOSP can calculate the support of non-overlapping sequential patterns with greater accuracy than an approximate calculation. It improves confidence in the mining of non-overlapping sequential patterns. Besides, users don't need to set $L$ in TNOSP.
	
To meet the above challenges, some improvements based on the Nettree are designed. The NETGAP algorithm is known to compute multiple times. After completing the NETGAP steps, the support for the corresponding $P$ in $S$ is calculated. Based on this, the steps are mainly divided into the following three parts:

\begin{itemize}	
    \item It creates an empty \textit{minheap} with a size of $k$ named \textit{Q}, which is sorted by the support of patterns and builds a candidate set named \textit{Candidate} to store candidate patterns. Then, it initializes \textit{minsup} to zero and \textit{Candidate} to patterns with a length of 1, which meet length constraints and gap constraints.
	
    \item It then iterates through all the existing candidates. If \textit{Q} is not full, it puts this candidate directly into \textit{Q}. If \textit{Q} is full, then it needs to compare the support of the candidate with the \textit{minsup}. If the support of the candidate is greater than \textit{minsup}, it pops the top element of \textit{Q} from the heap, and adds the candidate as a new element of \textit{Q}. In addition, it updates the value of \textit{minsup}.
	
    \item The next step is to generate the candidate set for the next length from the frequent patterns of the previous length. The above process is then repeated until no new candidates are generated. Finally, it outputs patterns in \textit{Q}.
\end{itemize}

\begin{algorithm}[h]
	\caption{TNOSP algorithm}
	\label{alg:top-k}
	\LinesNumbered
	\KwIn{sequence $S$; the number of top patterns \textit{k}; gap constraint \textit{gap} = [\textit{mingap, maxgap}] and length constraint \textit{len} = [\textit{minlen, maxlen}].} 
	\KwOut{Top-$k$ non-overlapping sequential patterns.}
	
	initialize \textit{Q} $\leftarrow$ $\varnothing$, \textit{minsup} $\leftarrow$ 0, $i$ $\leftarrow$ 1 and initialize \textit{FreArr} $\leftarrow$ $1$-patterns\; 
	initialize \textit{Candidate} $\leftarrow$ \textit{generateCandidate}(\textit{FreArr}, \textit{gap}, \textit{len})\;
	\While{\textit{Candidate} $\ne$ null \do}{
		\textit{FreArr} = $\varnothing$\;
		\For{$p$ $\in$ \textit{Candidate}}{
			
			initialize \textit{sup} $\leftarrow$ \textit{NETGAP}(\textit{S}, \textit{p}, \textit{minsup}, \textit{gap}, \textit{len})\;
			\If{\textit{sup} $>$ \textit{minsup} || (\textit{sup} == \textit{minsup} \&\& $Q$.\textit{size}() $<$ $k$)}{
				\textit{FreArr}.\textit{add}($p$)\;
				\If{$Q$.\textit{size}() $<$ $k$}{
					$Q$.\textit{add}($p$); \qquad $//$ Add mode $P$ to the minheap $Q$
					 \\
				}
				\ElseIf{$Q$.\textit{size}() == $k$}{
					update \textit{minsup}\;
					\If{\textit{sup} $>$ \textit{minsup}}{
						$Q$.\textit{update}($p$); \qquad $//$ Remove the mode less than the \textit{minsup} from the minheap $Q$ and add $p$ to the $Q$
					}	
				}
			}
		}
		remove from \textit{FreArr} those patterns with support less than \textit{minsup};\qquad $//$ the QMSP strategy\\
		\textit{Candidate} = \textit{generateCandidate}(\textit{FreArr}, \textit{gap}, \textit{len})\;
	}	
	\textbf{return} $Q$
\end{algorithm}

\subsection{The QMSP Strategy}  \label{QMSP}

Based on the above statement, the precision and credibility of the algorithm for mining non-overlapping sequential patterns have greatly improved. In addition, the pursuit of efficiency is also a part of the needs in algorithm optimization, so a pruning strategy named queue meta set pruning (QMSP) is proposed to improve efficiency. According to the previous subsection, the queue meta named \textit{freArr} in the Algorithm \ref{alg:top-k} of the length \textit{L}+1 is generated by the length of $L$ which is used to generate super-patterns. It causes some waste of resources to generate \textit{freArr} of the length \textit{L}+1 if the support of the pattern corresponding \textit{freArr} of the length \textit{L} is less than the \textit{minsup}. Because although the patterns that do not satisfy top-$k$ are no longer in the \textit{minheap} after the minheap is updated, the super-patterns of these patterns are still generated.
	
Hence, the QMSP strategy is proposed to reduce the memory cost to generate super-patterns and improve computing efficiency. The main idea is as follows: The candidate set of the next length $L$+1 is generated from the frequent patterns of the previous length $L$, as stated in the previous subsection. In the original NOSEP, the \textit{minsup} is fixed, so as long as the candidates have entered \textit{Q}, it will enter the queue meta \textit{freArr}. Therefore, candidates with the support that is less than the latest \textit{minsup} are removed after each round of $L$-patterns traversal. It will drastically reduce the number of candidate sets in the length $L$ + 1. 

In order to evaluate the efficiency of the QMSP algorithm, we compare the differences between visited nodes before and after using the QMSP strategy in Section \ref{Efficiency Evaluate} and Section \ref{VisNodes}. For example, the sequence $S$ = $s_1s_2s_3s_4s_5s_6s_7s_8s_9 = $ \textit{AGTCAGCAC}, where \textit{len} and \textit{gap} are [1,9] and [0,3], respectively. Besides, the parameter \textit{k} is set to 3. We then exemplify the TNOSP algorithm with the QMSP strategy below.

\begin{itemize}		
    \item Firstly, it creates a minheap for placing the three patterns whose support is the largest. The heap is empty at the start of the algorithm, and \textit{minsup} is set to 0.
		
    \item Secondly, it generates patterns with a length of 1, which meet length constraints and gap constraints, that is, \textit{A, C, G, T}. It is worth mentioning that the patterns are generated in lexicographical order, and their support is calculated by the NATGAP algorithm. Their supports are calculated and shown in Table \ref{Support}. Thus, the first generated pattern is \textit{A}, whose support is 3. It is the first one to be placed in the heap. Since the heap is non-full, the next patterns to enter the heap are \textit{C} and \textit{G}, whose supports are 3 and 2, respectively. At this point, the heap is full, and the value of the \textit{minsup} is updated to 2. The last pattern of length 1 is \textit{T}. Since its support is less than \textit{minsup} 2, it is rejected into the heap.
		
    \item Then, it generates super-patterns based on the previous step. According to the Apriori property and the QMSP strategy, pruning will be performed as follows: the patterns that do not enter the heap, that is, the patterns whose support is less than the current \textit{minsup} will be eliminated. Here, \textit{T} is no longer used to generate super-patterns. Only 9 patterns of length 2 will be generated, which are \textit{AA}, \textit{AC}, \textit{AG}, \textit{CA}, \textit{CC}, \textit{CG}, \textit{GA}, \textit{GG}, lexicographically. Comparing these 9 patterns of length 2 with the \textit{minsup} in order, it is obvious from Table \ref{Support} that pattern \textit{AC} is in the heap while pattern \textit{G} is popped off the heap. Since there is no pattern with greater support than \textit{minsup}, the iteration ends.
\end{itemize}

\begin{table}[h]
	\centering
	\caption{The support of all patterns with the lengths of 1 and 2 in $S$}
	\label{Support}
	\begin{tabular}{|c|c|c|c|c|c|c|c|c|c|c|c|c|c|c|c|c|}  
		\hline 
		\textbf{Patterns} & \multicolumn{4}{c|}{\textbf{A}} & \multicolumn{4}{c|}{\textbf{C}} & \multicolumn{4}{c|}{\textbf{G}} & \multicolumn{4}{c|}{\textbf{T}} \\
		\hline
		\textbf{Support} & \multicolumn{4}{c|}{3} & \multicolumn{4}{c|}{3} & \multicolumn{4}{c|}{2} & \multicolumn{4}{c|}{1} \\
		\hline
		\textbf{Patterns} & \textbf{AA} & \textbf{AC} & \textbf{AG} & \textbf{AT} & \textbf{CA} & \textbf{CC} & \textbf{CG} & \textbf{CT} & \textbf{GA} & \textbf{GC} & \textbf{GG} & \textbf{GT} & \textbf{TA} & \textbf{TC} & \textbf{TG} & \textbf{TT} \\
		\hline  
		\textbf{Support} & 2 & 3 & 2 & 1 & 2 & 2 & 1 & 0 & 2 & 2 & 1 & 1 & 1 & 1 & 1 & 0 \\	\hline	
	\end{tabular}
\end{table}

%% file: 5_experiment.tex
\section{Experiments}  \label{sec:experiments}

In the comparison, the precision of mining results needs to be judged. Precision is defined as the proportion of the discovered accurate patterns in all the mined patterns. In the approach that is proposed by Chai \textit{et al.} \cite{chai2018top}, the precision of the top-$k$ mining results is measured by the ratio calculation formula with a weight. In the following experiments, to facilitate calculation and statistics and re-establish consistent judgment standards, we revise the approach, and the equation can be stated as \textit{precision} = $\frac{\sum_{i=1}^{L}c_i}{\sum_{i=1}^{L} t_i}$. In this equation, $c_i$ and $t_i$ are the numbers of right top-$k$ frequent patterns and total top-$k$ frequent patterns with the length in $i$, respectively. And the length $L$ is stated in Section \ref{sec:preliminaries}. In general, $ \sum_{i=1}^{L}c_i$ = $\sum_{i=1}^{L}t_i $ = $k$ as long as $k$ is not less than the number of frequent patterns.

In order to assess the precision, performance, and scalability of the TNOSP algorithm we proposed, the experiments were carried out on a 64-bit, Windows 10 personal computer equipped with an Intel (R) Core™ i7-8750H CPU and RAM with 8 GB. In addition, TNOSP is programmed and implemented in Java, using Eclipse IDE as the development environment. The algorithm TNOSP was compared to the state-of-the-art NOSTOPK algorithm \cite{chai2018top}. Except for $k$, there are still some parameters that can affect the performance of the proposed algorithm, such as the length constraints and gap constraints. To ensure the consistency of other variables in the experiment, two constraints are set to \textit{len} $= [1, 20]$ and \textit{gap} $= [0, 5]$ in the following experiments. For ease of comparison, the output of the NOSTOPK was changed to all eligible patterns with lengths of 1 to $L$. In other words, outputs shorter than $L$ are no longer emptied before each iteration, so the top-$k$ patterns include all patterns with lengths up to $L$. We carried out some experiments on six real datasets. The details of these experiments are as follows.

\subsection{Experimental Datasets}

We conducted these experiments using six datasets to assess the efficiency and precision of the TNOSP algorithms. Three datasets are DNA sequence fragments of patients with certain diseases caused by viruses. These datasets can be obtained from the National Center for Biotechnology Information (NCBI)\footnote{\url{https://www.ncbi.nlm.nih.gov/labs/virus/vssi/}}. These three datasets all include several DNA sequences from different patients. These three datasets are called EBHF, MERS, and COVID-19, respectively. As the names of the datasets suggest, these three datasets each represent DNA fragments from multiple patients with Ebola hemorrhagic fever, Middle East Respiratory Syndrome, and Coronavirus Disease 2019. Each sequence represents the DNA sequence of a patient.

$ \bullet $ \textit{\textbf{EBHF}} is a collection of DNA sequence fragments from several patients suffering from Ebola hemorrhagic fever (EBHF) caused by the Ebola virus.

$ \bullet $ \textit{\textbf{MERS}} is a dataset that contains DNA sequence fragments from several patients with the Middle East Respiratory Syndrome (MERS) caused by the Middle East Respiratory Syndrome Coronavirus (MERS-CoV).

$ \bullet $ \textit{\textbf{COVID-19}} is still a dataset that contains DNA sequence fragments from several patients with the well-known Coronavirus Disease 2019 (COVID-19), caused by the novel coronavirus discovered in 2019 (2019-nCoV).

The other three datasets are also DNA sequence fragments but obtained from different animals. There are three datasets that are from DNA sequences of \textit{dog}, \textit{chimpanzee}, and \textit{human}, respectively. These datasets can be found at Kaggle\footnote{\url{https://www.kaggle.com/nageshsingh/dna-sequence-dataset}}. We also trimmed the sequences in the Chimp and human datasets because the experiments did not require such a large amount of data.

$ \bullet $ \textit{\textbf{dog}} is a dataset that contains DNA sequence fragments from a dog.

$ \bullet $ \textit{\textbf{chimpanzee}} is a dataset that contains DNA sequence fragments of a chimpanzee, the sequences of which are trimmed for the reason that the experiments don't need such a huge amount of data.

$ \bullet $ \textit{\textbf{human}} is still a dataset that are DNA sequence fragments of a human, the sequences of which are trimmed because the experiments don't need such a huge amount of data.

\begin{table}[h]
	\caption{Details of experimental datasets}
	\label{DATA}
	\centering
	\begin{tabular}{|c|c|c|c|}
		\hline
		\textbf{Sequence Dataset} & \textbf{Total Length} & \textbf{Number of Sequences} & \textbf{Average Length of each Sequence}  \\ \hline   \hline
		\textit{EBHF} & 118,779 & 6	& 19,796.5 \\ \hline
		\textit{MERS} & 181,126 & 6 & 30,187.7  \\ \hline
		\textit{COVID-19} & 536,097 & 18 & 29,783.2 \\ \hline
		\textit{dog} & 1,664,669 & 820	& 2,030.1 \\ \hline
		\textit{chimpanzee} & 1,711,934 & 873 & 1961.0 \\ \hline
		\textit{human} & 2,231,011 & 1,691 & 1,319.3 \\ \hline
	\end{tabular}
\end{table}

\subsection{Precision Evaluate}

The precision of TNOSP and NOSTOPK experiments with the six datasets mentioned above is measured in this subsection. NOSTOPK is a heuristic algorithm, while TNOSP is a precise algorithm. Therefore, a series of experiments are carried out to compare the precision of the two algorithms. In the experiments, we first test TNOSP and obtain the longest length in the set of output patterns, defined as $L_{max}$. The values of $L_{max}$ after testing are shown in Table \ref{Lmax}. Then assign $L_{max}$ to the $L$ in NOSTOPK. The results of the experiment are shown in Figure \ref{Precision}. It is shown that despite the high precision of NOSTOPK, there is still no way to complete all mining tasks with complete precision, which means that it is possible to miss some vital patterns. However, TNOSP can discover all patterns with precision, achieving 100\% in all mining tasks. Therefore, the above situation of pattern omission doesn't occur with TNOSP, which is of vital importance.

\begin{table}[h]
	\caption{$L_{max}$ of TNOSP with various $k$ in six datasets}
	\label{Lmax}
	\centering
	\begin{tabular}{|c|c|c|c|c|c|c|}
		\hline
		\textbf{Dataset} & \textbf{EBHF} & \textbf{MERS} & \textbf{COVID-19} & \textbf{dog} & \textbf{chimpanzee} & \textbf{human} \\ \hline
		$k$ = 10 & 5 & 5 & 5 & 3 & 3 & 3 \\ \hline
		$k$ = 20 & 6 & 6 & 6 & 5 & 4 & 4 \\ \hline
		$k$ = 30 & 7 & 7 & 6 & 5 & 5 & 5 \\ \hline
		$k$ = 40 & 7 & 7 & 7 & 5 & 5 & 5 \\ \hline
		$k$ = 50 & 7 & 7 & 7 & 6 & 6 & 5 \\ \hline
		$k$ = 60 & 7 & 7 & 7 & 6 & 6 & 6 \\ \hline
	\end{tabular}
\end{table}

\begin{table}[h]
	\caption{The precision with various $k$ in different datasets}
	\label{Precision}
	\centering
	\begin{tabular}{|c|c|c|c|c|c|c|c|}
		\hline
	    \textbf{dataset} & \textbf{Algorithm} & $k$ = 10 & $k$ = 20 & $k$ = 30 & $k$ = 40 & $k$ = 50 & $k$ = 60  \\ \hline
	    \multirow{2}{*}{\textbf{EBHF}}
	    &\textbf{TNOSP} & 100.0\% & 100.0\% & 100.0\% & 100.0\% & 100.0\% & 100.0\%  \\ \cline{2-8}
		&\textbf{NOSTOPK} & 100.0\% & 95.0\% & 93.3.0\% & 90.0\% & 92.0\% & 95.0\%  \\ \hline
		\multirow{2}{*}{\textbf{MERS}}
		&\textbf{TNOSP} & 100.0\% & 100.0\% & 100.0\% & 100.0\% & 100.0\% & 100.0\%  \\  \cline{2-8}
		&\textbf{NOSTOPK} & 100.0\% & 100.0\% & 100.0\% & 100.0\% & 100.0\% & 100.0\%  \\ \hline
		\multirow{2}{*}{\textbf{COVID-19}}
		&\textbf{TNOSP} & 100.0\% & 100.0\% & 100.0\% & 100.0\% & 100.0\% & 100.0\%  \\  \cline{2-8}
		&\textbf{NOSTOPK} & 100.0\% & 100.0\% & 100.0\% & 100.0\% & 100.0\% & 95.0\%  \\ \hline
		\multirow{2}{*}{\textbf{dog}}
		&\textbf{TNOSP} & 100.0\% & 100.0\% & 100.0\% & 100.0\% & 100.0\% & 100.0\%  \\  \cline{2-8}
		&\textbf{NOSTOPK} & 100.0\% & 95.0\% & 93.3\% & 95.0\% & 92.0\% & 95.0\%  \\ \hline
		\multirow{2}{*}{\textbf{chimpanzee}}
		&\textbf{TNOSP} & 100.0\% & 100.0\% & 100.0\% & 100.0\% & 100.0\% & 100.0\%  \\  \cline{2-8}
		&\textbf{NOSTOPK} & 100.0\% & 100.0\% & 93.3\% & 97.5\% & 94\% & 96.7\%  \\ \hline
		\multirow{2}{*}{\textbf{human}}
		&\textbf{TNOSP} & 100.0\% & 100.0\% & 100.0\% & 100.0\% & 100.0\% & 100.0\%  \\  \cline{2-8}
		&\textbf{NOSTOPK} & 100.0\% & 100.0\% & 97.5\% & 95.0\% & 96.0\% & 96.7\%  \\ \hline
	\end{tabular}
\end{table}

In general, the main idea of NOSTOPK is as follows. If a pattern $P_{sub}$ is a top-$k$ pattern with length $L$, there is a high probability that its super-pattern $P_{super}$ is a top-$k$ pattern with length ($L$ + 1), but it is not entirely precise. It should be noted that the top-$k$ referred to is for each length, but in the experiment, the output is unified as top-$k$ patterns in all the lengths. From a macro perspective, it is clear that the characteristics of the datasets have a great influence on precision. As shown in Table \ref{Precision}, the characteristics of the datasets MERS and COVID-19 are quite similar, and their precision is quite high. The precision of the MERS dataset, in particular, is 100\%. What's more, in the case of top-10 results, the precision can reach 100\% as well. However, it is shown in Figure \ref{Precision} that NOSTOPK is unstable in terms of precision. As $k$ increases, the precision is not monotonically decreasing or increasing but is not regular. Thus, as mentioned above, the precision of the algorithm depends to some extent on the dataset. Although the precision of NOSTOPK is already very high, reaching more than 90\%, as far as the algorithm TNOSP is concerned, users can use this algorithm to precisely mine all eligible non-overlapping sequence patterns, which runs pretty smoothly in terms of precision. Hence, we can conclude that TNOSP is a more advanced algorithm in terms of precision.

\subsection{Efficiency Evaluate} \label{Efficiency Evaluate}

\begin{figure}[h]
    \centering
    \includegraphics[clip,scale=0.45]{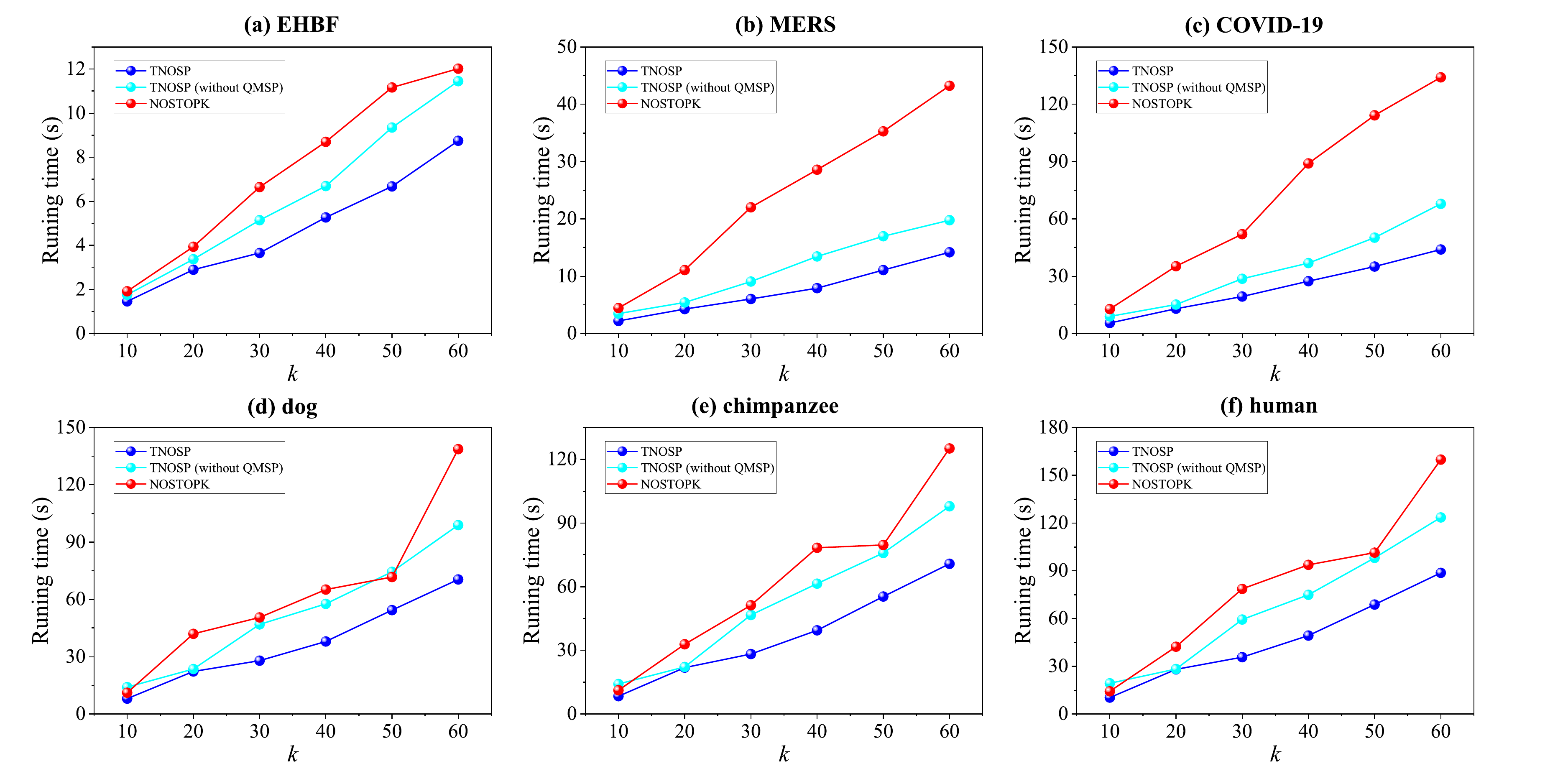}
    \caption{Running time with various $k$}
    \label{RunningTime}
\end{figure}

In this subsection, a series of experiments are performed to study the execution efficiency of TNOSP. Note that the running time consists of the runtime of the algorithm and file I/O processing time. Hence, in the experiments, the running time of the algorithm can weigh its efficiency, where ``running time" is defined as the time elapsing from the time that the program has just started running to the time that all eligible frequent sequential patterns have been found. Under this indicator, we experiment with TNOSP and NOSTOPK with the six datasets mentioned above. The difference and comparison of the running times of the two algorithms when $k$ is 10, 20, 30, 40, 50, and 60 on the DNA sequences are shown in Figure \ref{RunningTime}. In addition, in order to evaluate the efficiency of the QMSP strategy mentioned above, we also added the TNOSP algorithm without the QMSP strategy into the comparison.

As far as both algorithms are concerned, as $k$ increases, the size of the \textit{minheap} increases as well. The number of judgments to enter the heap and the number of calculated patterns also increase, as do $L_{max}$ in TNOSP and $L$ in NOSTOPK, increasing with running time for the reason that patterns of length $L+1$ are obtained from those of length $L$ according to the NETGAP algorithm. Based on the above, we make some comparisons between TNOSP and NOSTOPK. Figure \ref{RunningTime} shows that the running time of TNOSP is always shorter than that of NOSTOPK for all $k$ values and datasets, especially for MERS and COVID-19. This proves that TNOSP shows better performance and is more suitable for cases when the average length of each sequence is larger. In addition, the increase in running time of TNOSP is more balanced compared with NOSTOPK, especially in the datasets dog, chimpanzee, and human. In the case of NOSTOPK, Figure \ref{RunningTime} shows intuitively that some running times increase only slightly as $k$ increases, while others increase dramatically. To a certain extent, it also shows that TNOSP is not only more efficient in terms of running time but also has higher stability.

In addition, the pruning strategies of TNOSP will be analyzed as follows: In general, TNOSP without the QMSP strategy basically has a shorter running time than NOSTOPK (see Figure \ref{RunningTime}). However, there are still a few cases where the efficiency of TNOSP without the QMSP strategy is still low, which occurs in the last three datasets with $k$ = 10 and the dataset dog with $k$ = 50. However, it can be clearly seen that the efficiency of TNOSP after using QMSP is higher than that of NOSTOPK, and it is also significantly improved compared to the case without QMSP. It is enough to show that the QMSP strategy has a strong pruning effect, which has played a significant role in improving the efficiency of the TNOSP.

According to the experiment results, the TNOSP algorithm has better execution efficiency on all datasets when performing traditional top-$k$ mining tasks, which saves a significant amount of computation time when testing on large datasets. Following that, we will evaluate the algorithm's performance. In NOSTOPK, the purpose of using a heuristic algorithm is to improve efficiency significantly and provide results that have a small deviation. It is equivalent to sacrificing a certain degree of precision to obtain higher execution efficiency. It proves that in the case of using the Nettree, using the \textit{minheap} to adjust \textit{minsup} to deal with the top-$k$ problem and using QMSP to reduce the amount of computation both have great advantages. On the premise of precise mining, TNOSP can also achieve higher efficiency, indicating that the performance of the algorithm has been greatly improved.

\subsection{Evaluation of Visited Nodes} \label{VisNodes}

Because in the Nettree-based algorithm, the nodes need to be visited to get patterns and super-patterns of each eligible pattern. As a result, the visited nodes can be used as an important indicator to evaluate algorithm performance. We still carried out experiments, and the results of the experiments are shown in Figure \ref{VisitedNodes}. Similarly, as $k$ increases, the number of visited nodes increases as well. The visited nodes of TNOSP are all less than those of NOSTOPK under all $k$ values and datasets. And the increase in visited nodes in TNOSP is more balanced as well. In summary, the number of visited nodes corresponds to the running time of the previous subsection, which proves that the number of visited nodes is an important factor affecting computation time and execution efficiency.

\begin{figure}[h]
    \centering
    \includegraphics[clip,scale=0.46]{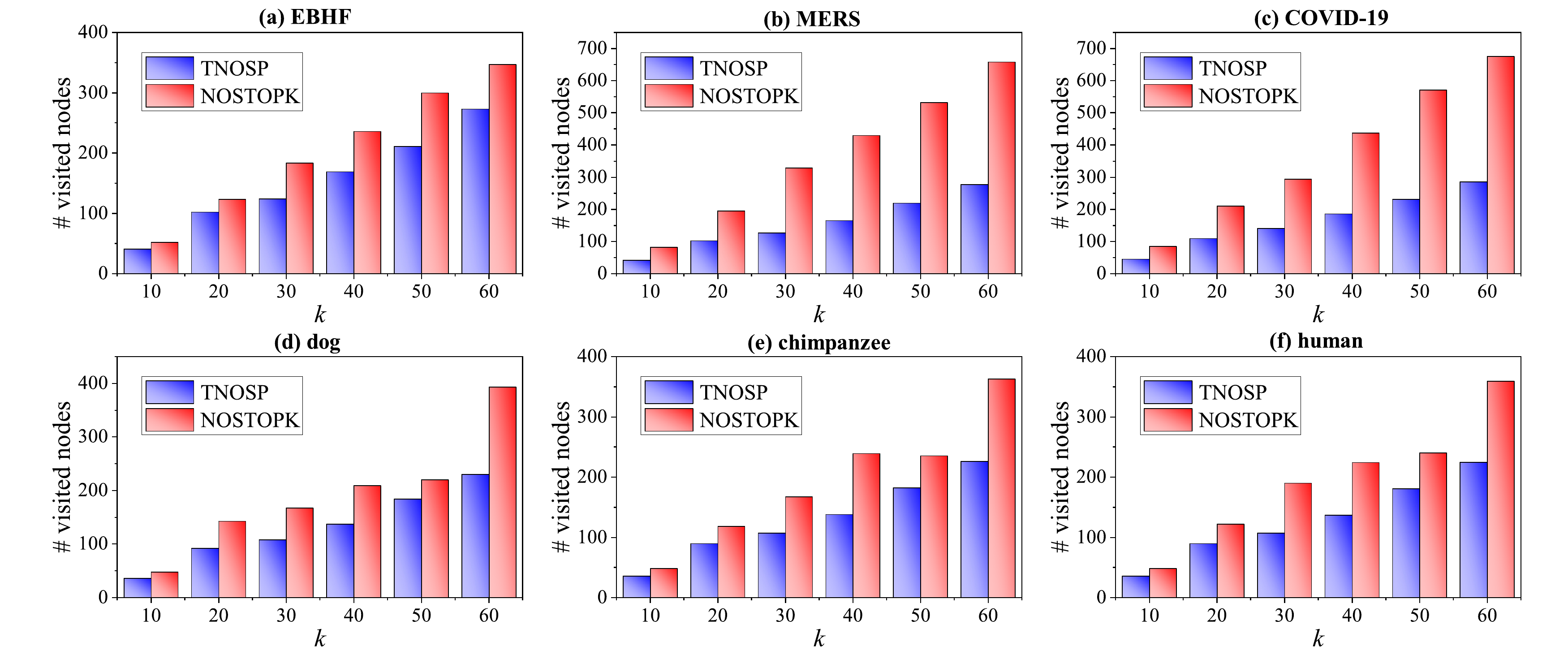}
    \caption{Visited nodes with various $k$ between TNOSP and NPSTOPK}
    \label{VisitedNodes}
\end{figure}

Besides, in order to evaluate the QMSP strategy, experiments to demonstrate the efficiency of the strategy are conducted as follows. We analyze TNOSP and the algorithm before using the QMSP strategy and record and compare the number of visited nodes before and after using the QMSP strategy. In the above comparisons, it can be seen that, except for the case of $k$ = 20 in the dataset human, under different $k$ values and different datasets, all numbers of visited nodes after using the strategy are less than those without using QMSP. It can further explain the necessity of using the QMSP strategy.

\begin{figure}[h]
    \centering
    \includegraphics[clip,scale=0.46]{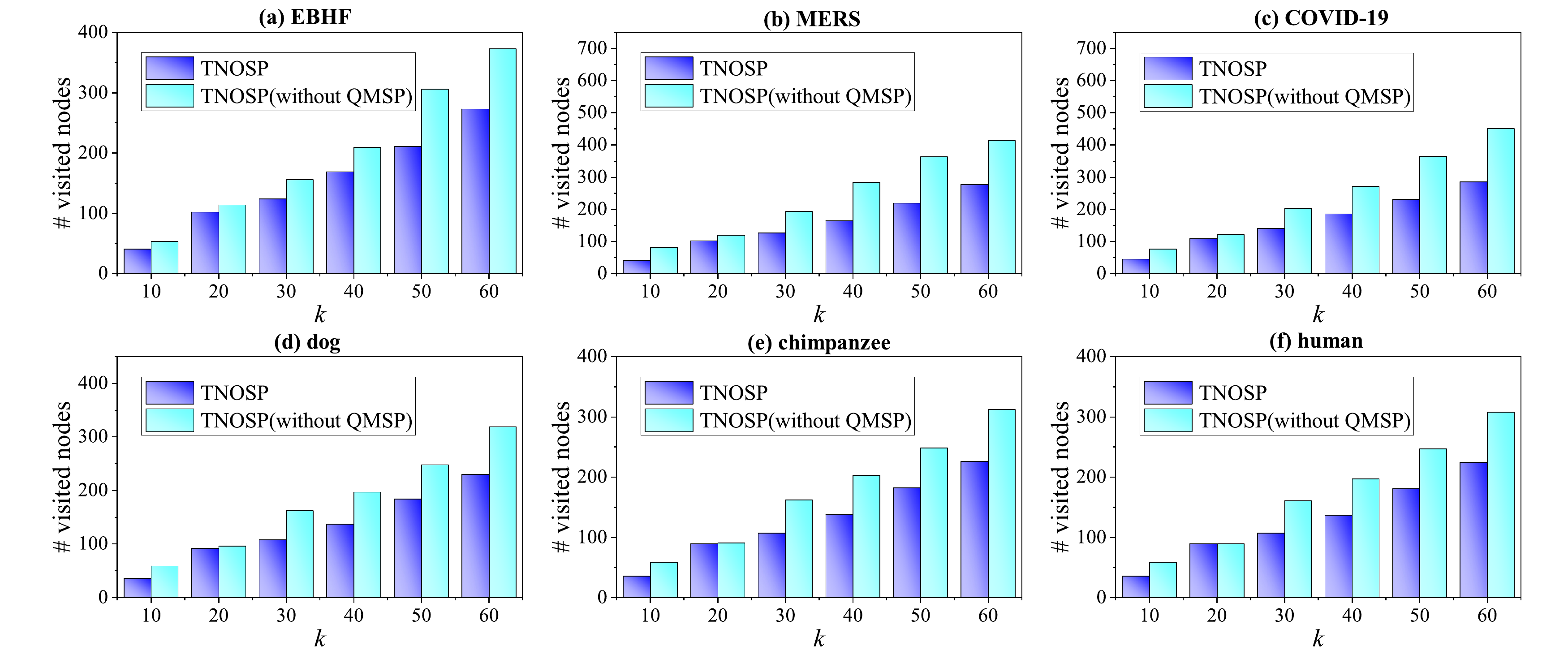}
    \caption{Visited nodes with various $k$ before and after using the QMSP strategy}
    \label{VisitedNodes}
\end{figure}

\subsection{Scalability}

For the purpose of measuring the scalability of TNOSP better, we carried out some experiments on eight datasets. All these datasets are the sub-datasets of the human dataset mentioned above, and the first 200 sequences of the human dataset are taken first for the following experiments. The size is then increased in 200-unit increments until the sequences reach 1,600. Therefore, their features are consistent with the human dataset in the experiment above.

\begin{figure}[h]
    \centering
    \includegraphics[clip,scale=0.18]{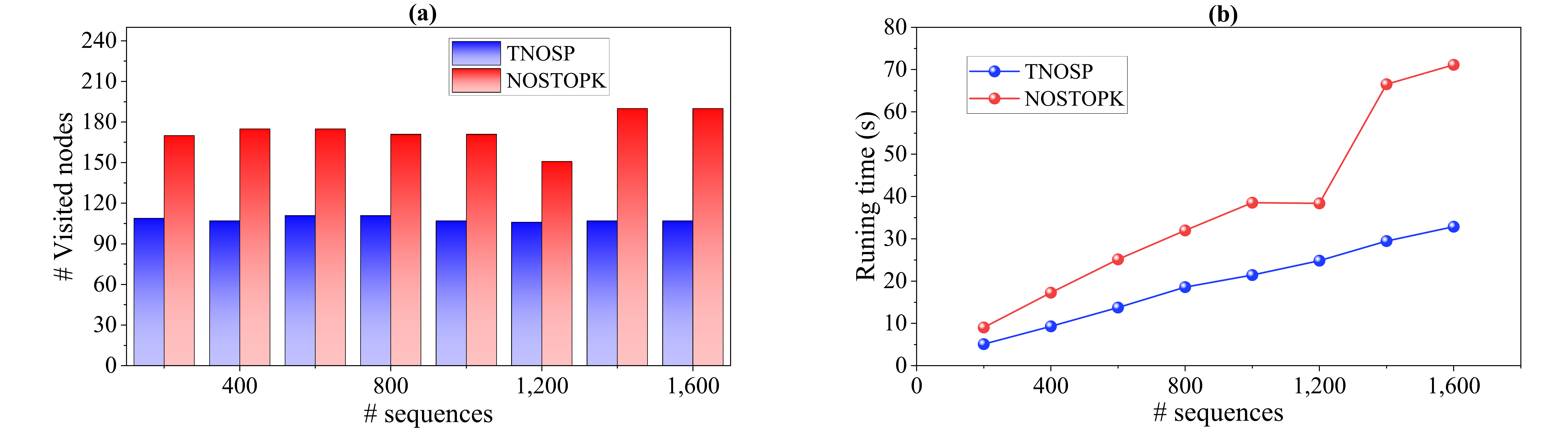}
    \caption{Scalability of the compared approaches when $k$ = 30 in a dataset}
    \label{fig_scalability}
\end{figure}

On the basis of running time and visited nodes, the results of experiments are shown in Figure \ref{fig_scalability}, which shows that the running time of TNOSP increases more slowly than that of NOSTOPK. And it can be noticed that the running time of NOSTOPK when the size of the sequences is 1,000 is almost the same as when the size is 800. The reason for this is that when the size is 1000, it's $L_{max}$ is 4, while $L_{max}$ is 5 when the size is another value. Obviously, as the dataset size increases, the running time of TNOSP shows a nearly linear growth relationship. What’s more, with the increase in the size of the datasets, the number of visited nodes in TNOSP is almost constant, while that in NOSTOPK changes. It means that the scalability of TNOSP on access nodes is stable. As shown in Figure \ref{fig_scalability}, as the size of the dataset increases, the running time of the program shows a linear relationship and will not grow too fast, and visited nodes present a stable state. Hence, we can draw the conclusion that the TNOSP algorithm is more scalable on long sequences and large datasets.

We further analyze the advantages of the algorithm. The above experiments demonstrate that TNOSP exhibits more advanced advantages when performing traditional top-$k$ non-overlapping sequential pattern mining. The advantages are manifested in various aspects, and they are shown directly through three quantifiable indicators: precision, running time, visited nodes, and scalability. The experimental results above show that TNOSP has higher stability and performs better in terms of precision, efficiency, and scalability compared with NOSTOPK. Hence, it can be concluded that TNOSP is a more advanced algorithm for mining traditional top-$k$ non-overlapping sequential patterns.

%% file: 6_conclusion.tex
\section{Conclusions and Future Works}  \label{sec:conclusion}
	
In this paper, to solve the problem that existing non-overlapping sequential pattern mining algorithms can not be focused on the most valuable and precise patterns, a new algorithm named TNOSP is proposed to discover top-$k$ non-overlapping sequential patterns without setting the minimum support threshold. TNOSP is more flexible and can more effectively search all eligible patterns. In addition, TNOSP doesn't need to specify the fixed length for desired patterns, which is more suitable than traditional top-$k$ mining algorithms and has more real-world applications. Finally, we conduct several experiments on six different datasets with different values of $k$. The experimental results show that TNOSP always has better precision, efficiency, and scalability. In future work, we plan to integrate TNOSP with some technologies, such as parallel processing, streaming data processing, and so on, to improve operating speed and reduce data or file Input/Output (I/O) processing time. We may investigate and propose a better method for mining top-$k$ patterns with non-overlapping sequences.

\section*{Acknowledgment}

This research was supported in part by the National Natural Science Foundation of China (Nos. 62272196 and 62002136), Natural Science Foundation of Guangdong Province (No. 2022A1515011861), and the Young Scholar Program of Pazhou Lab (No. PZL2021KF0023).